\documentclass[3p,times]{elsarticle}

\usepackage{bm}
\usepackage{amsmath}

\usepackage{ecrc}

\usepackage{epstopdf}

\volume{00}

\firstpage{1}

\journalname{Nuclear Physics A}

\runauth{T. Hirano et al.}

\jid{nupha}

\jnltitlelogo{Nuclear Physics A}

\usepackage{graphicx}
\usepackage{amsmath,amssymb}

\begin{document}

\begin{frontmatter}



\title{Multiplicity fluctuation from hydrodynamic noise}

\author[label1]{T. Hirano}
\author[label1,label2,label3]{R. Kurita}
\author[label1,label2,label3]{K. Murase}
\author[label1]{K. Nagai}

\address[label1]{Department of Physics, Sophia University, Tokyo 102-8554, Japan}
\address[label2]{Department of Physics, The University of Tokyo, Tokyo 113-0033, Japan}
\address[label3]{Theoretical Research Division, Nishina Center, RIKEN, Wako 351-0198, Japan}

\begin{abstract}
We discuss multiplicity fluctuation
caused by noises during hydrodynamic evolution of the quark-gluon fluid
created in high-energy nuclear collisions. 
\end{abstract}


\end{frontmatter}



\section{Introduction}
\label{intro}

Event-by-event analysis of high-energy 
nuclear collisions has been performed to understand 
 transport properties of the quark gluon plasma (QGP).
Higher order harmonics of collective flow
is intimately related with initial state fluctuation
in nucleus-nucleus (A+A) collisions \cite{Hirano:2012kj}.
In addition to this,
the system created in
high multiplicity proton-nucleus (p+A) events
also exhibits collective behaviour \cite{CMS:2012qk,Abelev:2012ola}.
These can be described by conventional
hydrodynamic simulations \cite{Bozek:2013ska}.
However, the apparent success of hydrodynamics
in such small systems has been questioned since the hydrodynamics
poses on a condition of gradually changing thermodynamic variables
in both time and space direction
for its validity.

Motived by these facts, we extended
the conventional hydrodynamic framework
by introducing causal hydrodynamic fluctuations \cite{Murase:2013tma}.
Within framework of the relativistic fluctuating hydrodynamics,
we discuss multiplicity fluctuation
caused by noises during hydrodynamic evolution of the quark-gluon fluid
created in high-energy nuclear collisions. 
We first discuss the stochastic equations for dissipative currents
coupled with the temporal evolution equation for the energy density
in one-dimensionally expanding coordinate system \cite{Bjorken:1982qr}
to demonstrate that the final entropy fluctuates from event to event for a given initial condition.
We next discuss the fluctuation theorem in non-equilibrium statistical mechanics \cite{GallavottiCohen,EvansSearles}.
We finally discuss consequences of hydrodynamic fluctuations on final multiplicity.

\section{Fluctuating hydrodynamics in Bjorken coordinates}
\label{sec:FH}

To demonstrate the time evolution of 
the QGP in high-energy nuclear collisions 
under hydrodynamic fluctuations,
we introduce an one-dimensionally expanding coordinate system 
and assume boost invariant scaling ansatz \cite{Bjorken:1982qr}.
In Bjorken coordinates 
$(\tau = \sqrt{t^2-z^2}, x, y, \eta_{s} = \frac{1}{2}\ln[(t+z)/(t-z)])$,
relativistic hydrodynamic equations reduce to \cite{Hirano:2008hy}
\begin{eqnarray}
\label{eq:eom-Bj}
\frac{de}{d\tau} & = & -\frac{e+P_{s}}{\tau} \left(1-\frac{\pi}{s T}+\frac{\Pi}{s T}   \right),\\
\pi & = & \pi^{tt}-\pi^{zz},
\end{eqnarray}
where $e$ is energy density, $P_{s}$ is hydrostatic pressure,
$T$ is temperature,
$s = (e+P_{s})/T$ is entropy density,  $\Pi$ is bulk pressure and
$\pi^{\mu \nu}$ is shear stress tensor.
In the case of perfect fluids, \textit{i.e.} $\Pi = \pi =0$, the entropy $S = s \tau \Delta \eta_{s} \Delta x \Delta y$
in a fluid element
is conserved.
In the case of dissipative fluids,
it is easy to show
\begin{eqnarray}
\frac{1}{\Delta\eta_{s} \Delta x \Delta y}\frac{dS}{d\tau} & = & s + \tau \frac{ds}{d\tau} = s+\frac{\tau}{T} \frac{de}{d\tau} =  s+\frac{\tau}{T}  \left(-\frac{e+P_{s}}{\tau} \right) \left(1 -\frac{\pi}{s T}+\frac{\Pi}{s T}  \right) \nonumber\\
& = &  \frac{\pi}{T}-\frac{\Pi}{T} .
\end{eqnarray}
Thus the production rate of the entropy in one fluid element becomes
\begin{equation}
\label{eq:entropyprod}
\sigma = \frac{dS}{d\tau} = \left( \frac{\pi}{T}-\frac{\Pi}{T}  \right)\Delta\eta_{s} \Delta x \Delta y
\end{equation}
In relativistic fluctuating hydrodynamics \cite{Murase:2013tma}, the constitutive equations in the differential form are
\begin{eqnarray}
\label{eq:eom_shear}
\tau_{\pi}\frac{d\pi}{d\tau} + \pi & = & \frac{4\eta}{3\tau} + \xi_\pi\\
\label{eq:eom_bulk}
\tau_{\Pi}\frac{d\Pi}{d\tau} + \Pi & = & -\frac{\zeta}{ \tau} + \xi_\Pi
\end{eqnarray}
Here $\tau_\pi$ and $\tau_\Pi$ are relaxation times,
$\eta$ and $\zeta$ are shear and bulk viscosities and  $\xi_\pi$ and $\xi_\Pi$ are Gaussian white noises for 
shear and bulk pressure, respectively.
It is worthwhile mentioning that the right hand side of 
Eq.~(\ref{eq:entropyprod}) is \textit{not} positive definite due to
the existence of noises in Eqs.~(\ref{eq:eom_shear})  and (\ref{eq:eom_bulk}).
In the conventional hydrodynamic framework,
constitutive equations are so designed to 
obey the second law of thermodynamics. On the other hand, 
stochastic constitutive equations are obtained
in fluctuating hydrodynamic framework
so that solutions of these equations obey
 the fluctuation-dissipation relation by taking an ensemble average.

\section{Fluctuation theorem}
\label{sec:FT}

After the linear response theory was established \cite{Kubo},
a variety of progress has been made  
 in non-equilibrium statistical mechanics.
Among them, the fluctuation theorem \cite{GallavottiCohen,EvansSearles}
has become a milestone in that field. 
Since the fluctuation theorem contains the Green-Kubo formula
at long-time limit, it is believed to capture 
some important properties of non-equilibrium processes
away from equilibrium. As seen in the previous section,
entropy production can be negative in a certain sample event.
Interestingly,  the probability for the system
having negative entropy production can be quantified through
the fluctuation theorem shown below.

$\sigma(t)$ is supposed to be the production rate of entropy and 
\begin{equation}
\bar{\sigma}(t) = \frac{1}{t} \int_{0}^{t} dt' \sigma(t')
\end{equation}
is its average over time duration $t$.
A relation between
a probability of $\bar{\sigma} = \alpha$ and 
the one of  $\bar{\sigma} = -\alpha$ holds as
\begin{eqnarray}
\label{eq:TFT}
\frac{P(\bar{\sigma} = \alpha)}{P(\bar{\sigma} = -\alpha)} & = & \exp(\alpha t),\\
\label{eq:SSFT}
\lim_{t\rightarrow \infty}\frac{1}{t}\ln 
\frac{P(\bar{\sigma} = \alpha)}{P(\bar{\sigma} = -\alpha)} & = & \alpha.
\end{eqnarray}
Equations~(\ref{eq:TFT}) and (\ref{eq:SSFT})
 are called ``transient fluctuation theorem'' and
``steady state fluctuation theorem'', respectively.
For details, see Refs.~\cite{GallavottiCohen,EvansSearles}.

Suppose entropy production rate obeys Gaussian
\begin{equation}
P(\bar{\sigma}) = \frac{1}{\sqrt{2 \pi a^2}}
\exp\left[ -\frac{(\bar{\sigma}- \langle \bar{\sigma}\rangle)^2}{2 a^2}\right],
\end{equation}
the transient fluctuation theorem (\ref{eq:TFT})
leads to $2\langle\bar{\sigma} \rangle/a^2 = t$.
Here $\langle \cdots \rangle$ denotes an ensemble average.
Thus entropy distribution at time $t$ is
\begin{equation}
P(S) \propto 
\exp\left[ -\frac{(S- \langle S \rangle)^2}{2 t^2 a^2}\right] = 
\exp\left[ -\frac{(S- \langle S \rangle)^2}{2 (\sqrt{2 \langle\bar{\sigma} \rangle t})^2}\right],
\end{equation}
So far, the above results
have been obtained in general under the fluctuation theorem.


\section{Entropy fluctuation}
\label{sec:multiplicity}

Now we focus on the  Bjorken expansion case discussed in Sec.~\ref{sec:FH}.
Relative width of
entropy distribution at time $\tau$ 
(being sufficiently larger than the relaxation time $\tau_{\pi/\Pi}$) becomes
\begin{equation}
\frac{\Delta S}{\langle S \rangle}(\tau) = 
\frac{\sqrt{2\langle \Delta (\tau s)\rangle}}{\tau_0 s_0 + \langle \Delta (\tau s)\rangle} \frac{1}{\sqrt{\Delta \eta_{s} \Delta x \Delta y}}.
\end{equation}
Here $\tau_0$ is initial time, $s_0$ is initial entropy density and
$\langle \Delta (\tau s)\rangle$ is average entropy production 
per volume of 
local thermal system $\Delta V = \tau \Delta \eta_{s} \Delta x \Delta y$.
In heavy ion collisions, the number of independent
local thermal system in the transverse plane
can be estimated as $N = A(b)/ \Delta x \Delta y$.
Here $A(b)$ is the effective transverse area of collision geometry
for a given impact parameter $b$.
Finally relative fluctuation for total entropy per  
space-time rapidity window $\Delta \eta_{s}$ becomes
\begin{eqnarray}
\label{eq:entropyfluctuation}
\frac{\Delta S_{\mathrm{tot}}}{\langle S_{\mathrm{tot}}\rangle}
& = & \frac{1}{\sqrt{N}} \frac{\Delta S}{\langle S \rangle}= 
\frac{\sqrt{2\langle \Delta (\tau s)\rangle}}{\tau_0 s_0 + \langle \Delta (\tau s)\rangle} \frac{1}{\sqrt{\Delta \eta_{s} A(b)}} \nonumber \\
& \leq & \frac{1}{\sqrt{2 \tau_0 s_0}} \frac{1}{\sqrt{\Delta \eta_{s} A(b)}}
= \frac{1}{\sqrt{2 S_{\mathrm{ini}}}}.
\end{eqnarray}
From the first line to the second line in Eq.~(\ref{eq:entropyfluctuation}),
we utilise an inequality $\sqrt{2}x/(x^2 + c^2 ) \leq 1/\sqrt{2}c$
for $x = \sqrt{\langle \Delta (\tau s)\rangle}$ and $c = \sqrt{\tau_0 s_0}$.
This result suggests 
fluctuation of total entropy 
is bounded by a value evaluated solely from  the initial entropy.
\begin{figure}
\begin{center}
\includegraphics*[width=9.cm, trim = 6pt 6pt 6pt 6pt]{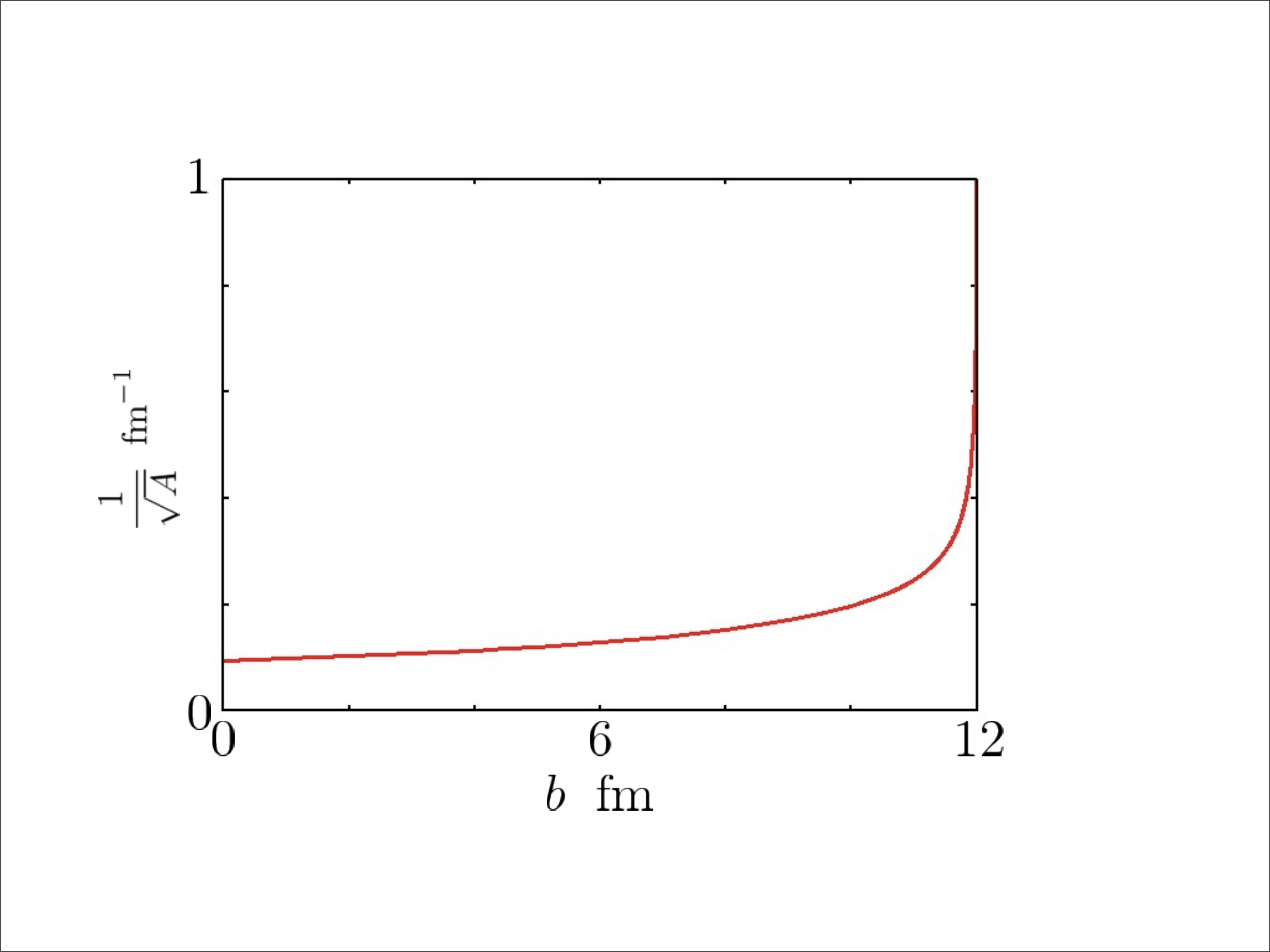}\\
\caption{Impact parameter dependence of
a factor $1/\sqrt{A}$ which controls the entropy fluctuation.}
\label{fig:cent_dep}
\end{center}
\end{figure} 

It turns out that entropy fluctuation per 
rapidity window $\Delta \eta_{s}$ depends on a factor 
$1/\sqrt{\Delta \eta_{s}A}$.
Figure \ref{fig:cent_dep}
shows $1/\sqrt{A}$ as a function of impact parameter $b$
by assuming $A$ corresponds to overlap area of collisions
of two nuclei with a radius of 6 fm.
The factor $1/\sqrt{A}$ 
gradually increases with $b$ but rapidly enhances
in peripheral collisions $b \sim 10$-$12$ fm.
Identifying inelastic cross section of p+p collisions at RHIC and the LHC
with the above transverse area $A$,
$1/\sqrt{\sigma_{\mathrm{in}}} \approx 0.4$-$0.5$ fm$^{-1}$.
These values are comparable with the ones in peripheral A+A collisions.
As expected, the effects of entropy fluctuation 
manifest in small system.

\section{Summary}

Entropy and, in turn, multiplicity fluctuate
from event to event 
due to hydrodynamic fluctuation of dissipative
currents such as shear stress tensor and bulk pressure
even if the initial state is the same in a macroscopic sense.
Although \textit{event-averaged} entropy has to increase with time
so that the system obeys the second law of thermodynamics,
entropy in a certain event can, however, 
\textit{decrease} with time temporarily and locally due
to the hydrodynamic fluctuations. 
The probability of decreasing entropy during hydrodynamic evolution
is tiny in general. 
Nevertheless, the probability is quantified by the fluctuation theorem
as known in the non-equilibrium statistical mechanics. 
The multiplicity fluctuation caused by hydrodynamic noise
must play a crucial role in small system
such as in peripheral A+A and/or 
in high multiplicity p+A and p+p events.

\section*{Acknowledgement}
The work was supported by JSPS KAKENHI Grant Numbers
25400269 (T.H.) and 12J08554 (K.M.).
The work of R.K. and K.M. was supported by an Advanced
Leading Graduate Course for Photon Science grant
in the University of Tokyo.
The work of K.M. was supported by a JSPS Research Fellowship
 for Young Scientists.








\end{document}